\documentclass[final,1p,times]{elsarticle} 

\usepackage{graphicx}
\usepackage{amssymb} 
\usepackage{amsthm} 
\usepackage{lineno}
\usepackage{units}
\usepackage{caption}

\journal{Nuclear Physics A} 

\begin{document}

\begin{frontmatter} 

\title{Measurement of Direct Photons in pp and Pb-Pb Collisions with ALICE}

\author{Martin Wilde (for the ALICE\fnref{col1} Collaboration)}
\fntext[col1] {A list of members of the ALICE Collaboration and acknowledgements can be found at the end of this issue.}
\address{Institut f\"ur Kernphysik, Westf\"alische Wilhelms-Universit\"at M\"unster, M\"unster, Germany}


\begin{abstract} 
The measurement of the direct photon transverse momentum spectrum in Pb-Pb collisions at \unit[$\sqrt{s_{_{NN}}}=2.76]{TeV}$ with data taken by the ALICE experiment is presented. The measurement shows a clear direct-photon signal for 0-40\% most central collisions below \unit[4]{GeV/$\mathrm{c}$} that can not be described by next-to-leading-order perturbative QCD (NLO pQCD) calculations. Above this value of $p_{\mathrm{T}}$ the result is in agreement with pQCD predictions. The low $p_{\mathrm{T}}$ signal is expected to have thermal photon contributions. The inverse slope parameter of an exponential fit is extracted as $T_{\mathrm{LHC}}=\unit[304\pm 51^{\mathrm{syst+stat}}]{MeV}$. For a baseline measurement the analysis is performed for proton-proton collisions at \unit[$\sqrt{s}=7]{TeV}$ and for peripheral (40-80\%) Pb-Pb collisions. Both results show no low $p_\mathrm{T}$ direct-photon signal and are in agreement with pQCD calculations.

\end{abstract} 
\end{frontmatter}
\section{Introduction}
Direct-photon production in hadronic collisions can be understood as a superposition of different production sources. In pp collisions, leading order pQCD processes, quark-gluon compton-scattering and quark-anti-quark annihilation are the main components of direct-photon production at high $p_{\mathrm{T}}$. The second source is photons produced in yet fragmentation \cite{theorypp}. In heavy-ion collisions these prompt photons should be enhanced by the larger number of binary collisions. They may be modified by isospin effects and nuclear shadowing. Fragmentation photons are affected by jet quenching; saturation effects may lead to further modifications for both sources \cite{theoryPbPbpQCD1,theoryPbPbpQCD2}.
 
In addition, final-state production mechanisms have to be considered in heavy-ion collisions. In the QGP phase, photons should be produced by the scattering of hard partons traversing the medium with thermalized partons, as well as by the scattering of thermalized partons. In the hot hadronic gas, produced after hadronization, direct photons should emerge from the scattering of the thermalized hadrons within the gas \cite{theoryPbPbpQCD1,theoryPbPbpQCD2,theoryPbPbpQCD3}.  The thermalized nature of the production medium (QGP and hadron gas) should be reflected in the $p_{\mathrm{T}}$ distribution of the produced thermal photons.

The measurement of direct photons in pp collision is an important test for the validation of pQCD. In heavy-ion collisions a thermal photon signal is expected. From this signal an average temperature can be extracted. The comparison to pp pQCD results at higher $p_{\mathrm{T}}$ tests the binary scaling behavior of initial hard scatterings.

\section{Analysis Method}
For both analyzed systems (Pb-Pb at $\sqrt{s_{_{NN}}} = \unit[2.76]{TeV}$, pp at $\sqrt{s} = \unit[7]{TeV}$) the same analysis method is used. The Pb-Pb analysis is divided into two bins of centrality (central: 0-40\%, peripheral: 40-80\%). The direct-photon signal is extracted via the subtraction method (see e.g. \cite{substraction}):
\begin{equation}\label{eq:subs}
  \gamma_{\mathrm{direct}} = \gamma_{\mathrm{inc}}-\gamma_{\mathrm{decay}} = (1-\frac{\gamma_{\mathrm{decay}}}{\gamma_{\mathrm{inc}}})\cdot \gamma_{\mathrm{inc}}.
\end{equation}
The method is based on the measurement of the inclusive photon yield via the reconstruction of their conversion products.  For this purpose a secondary vertex finder is used. The algorithm combines oppositely charged tracks with a large impact parameter. If a pair is accepted the conversion point and the momentum of the photon is calculated. The algorithm is generally used to identify weakly decaying particles like $\Lambda$ or $K_s^0$.

 To optimize the signal to background ratio and to exclude other sources of secondary vertices beside photons (combinatorial background, $\Lambda$, $K_s^0$, etc.), several photon selection criteria comparable to the criteria used in \cite{ppPi0Paper} are applied.

 The extracted raw photon spectrum is corrected for purity, reconstruction efficiency and conversion probability of photons in the detector estimated from MC simulations.

The decay photon spectrum, $\gamma_{\mathrm{decay}}$, is obtained by a cocktail calculation. The calculation is based on yield parametrizations of mesons with photon decay branches.
 The main source of decay photons ($\sim 80\%$) is $\pi^0\rightarrow\gamma\gamma$. The second largest contribution ($\sim 18\%$) is the decay of the $\eta$ meson ($\eta\rightarrow\gamma\gamma$). In the analyzed pp collisions both meson yields, $\pi^0$ and $\eta$, are measured \cite{ppPi0Paper}. In Pb-Pb only the $\pi^0$ yield is known \cite{QM2012}. In both cases additional sources of decay photons from unknown meson yields ($\eta$, $\eta^{\prime}$, $\omega$, $\phi$ and $\rho_0$) are obtained from $m_T$-scaling \cite{phenixscaling}.

To minimize the systematic uncertainties the ratio $\frac{\gamma_{\mathrm{decay}}}{\gamma_{\mathrm{inc}}}$ used in formula (\ref{eq:subs}) is calculated via:
\begin{equation}
  \label{eq:doubleratio}
  \frac{\gamma_{\mathrm{inc}}}{\gamma_{\mathrm{decay}}} = \frac{\gamma_{\mathrm{inc}}}{\pi^0}/\frac{\gamma_{\mathrm{decay}}}{\pi^0_{\mathrm{param}}}.
\end{equation}
The method guarantees an exact cancellation of all normalization factors used in the spectra. Further uncertainties are canceled due to the fact that the $\pi^0$ extraction is based on the same set of photon candidates as the inclusive photon yield \cite{substraction,substraction2}.

\section{Results}
\begin{figure}[h!]
\begin{minipage}[h]{0.5\textwidth}

\includegraphics[width=0.96\textwidth]{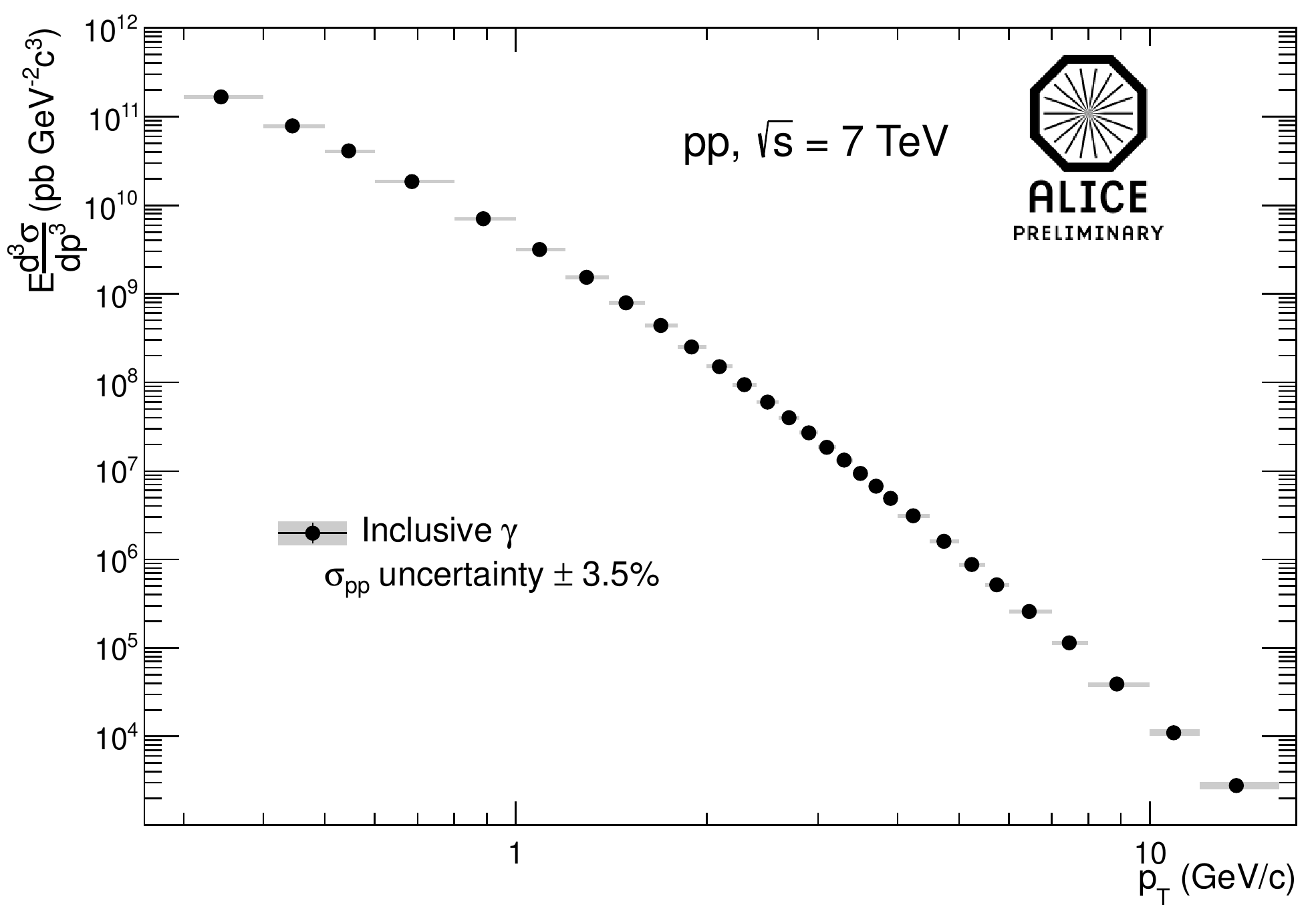}
\captionsetup{margin={0pt,5pt}}
\caption{Invariant cross section of inclusive photons in pp collisions at $\sqrt{s} = \unit[7]{TeV}$\newline}\label{incphotonpp}

\end{minipage}\hfill
\begin{minipage}[h]{0.5\textwidth}

\includegraphics[width=0.96\textwidth]{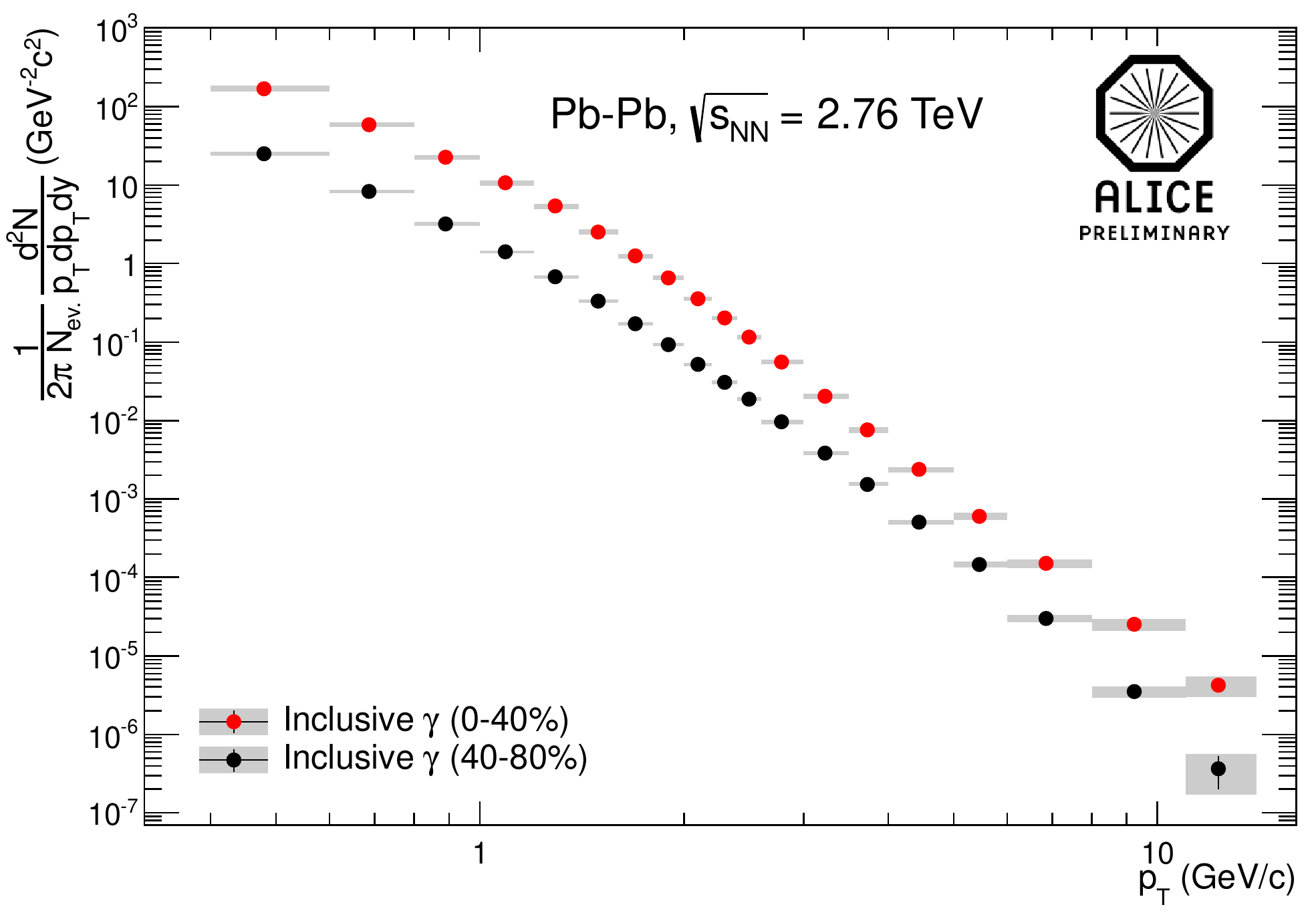}
\captionsetup{margin={5pt,0pt}}
\caption{(color online) Invariant yield of inclusive photons in Pb-Pb collisions at $\sqrt{s_{_{NN}}}=\unit[2.76]{TeV}$ for 0-40\% and 40-80\% centrality}\label{incphotonPbPb}
\end{minipage}\hfill
\end{figure}

Figure\,\ref{incphotonpp} shows the invariant cross section of inclusive photons produced in pp collisions at $\sqrt{s} = \unit[7]{TeV}$. It is calculated as:
\begin{equation}
  \label{eq:crosssection}
  E\frac{\mathrm{d}^3\sigma}{\mathrm{d}p^3}=\frac{1}{2\pi}\frac{\sigma_{MB_{OR}}}{N_{\mathrm{events}}}\frac{1}{p_{\mathrm{T}}}\frac{\mathcal{P}}{\mathcal{C}\mathcal{E}}\frac{N^{\gamma}}{\Delta y \Delta p_{\mathrm{T}}}.
\end{equation}

In the formula, $\mathcal{P}$, $\mathcal{C}$ and $\mathcal{E}$ are the purity, the conversion probability and the efficiency. They are obtained from MC. $\sigma_{MB_{OR}}$ is the interaction cross section for the MBOR trigger that is used for the pp collisions ($\sigma_{MB_{OR}} = \unit[62.2\pm 2.2]{mb}$ (syst)). $N_{\mathrm{events}}$ is the number of MBOR events that were collected and $N^{\gamma}$ represents the number of  photon candidates. In Fig.\,\ref{incphotonPbPb}, the invariant inclusive photon yields for both centralities in Pb-Pb are presented.

\begin{figure}[h!]
\begin{minipage}[h]{0.5\textwidth}

\includegraphics[width=0.96\textwidth]{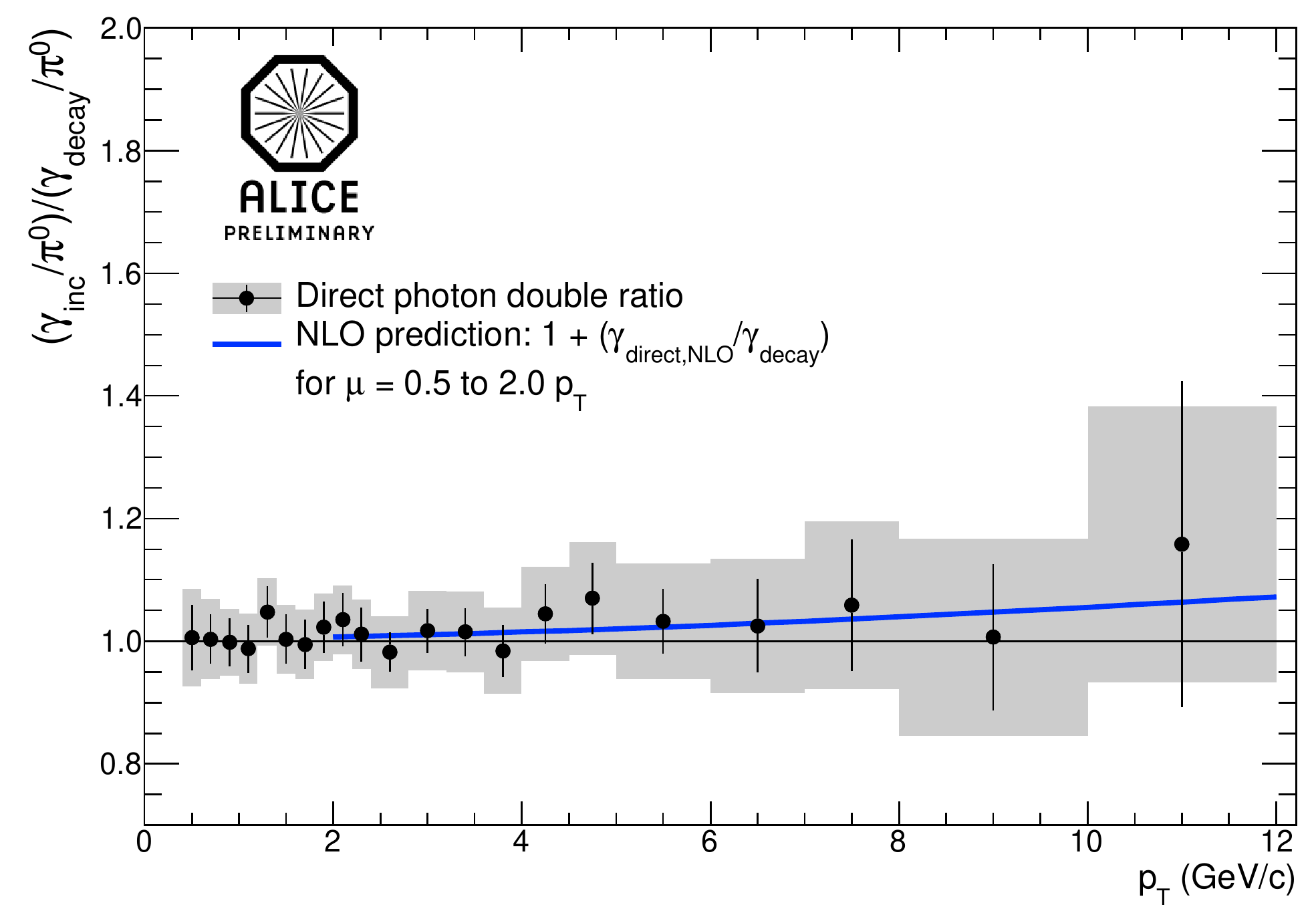}
\captionsetup{margin={0pt,5pt}}
\caption{(color online) Direct-photon double ratio in pp collisions at $\sqrt{s} = \unit[7]{TeV}$ with NLO pQCD predictions\newline}\label{doublepp}

\end{minipage}\hfill
\begin{minipage}[h]{0.5\textwidth}

\includegraphics[width=0.96\textwidth]{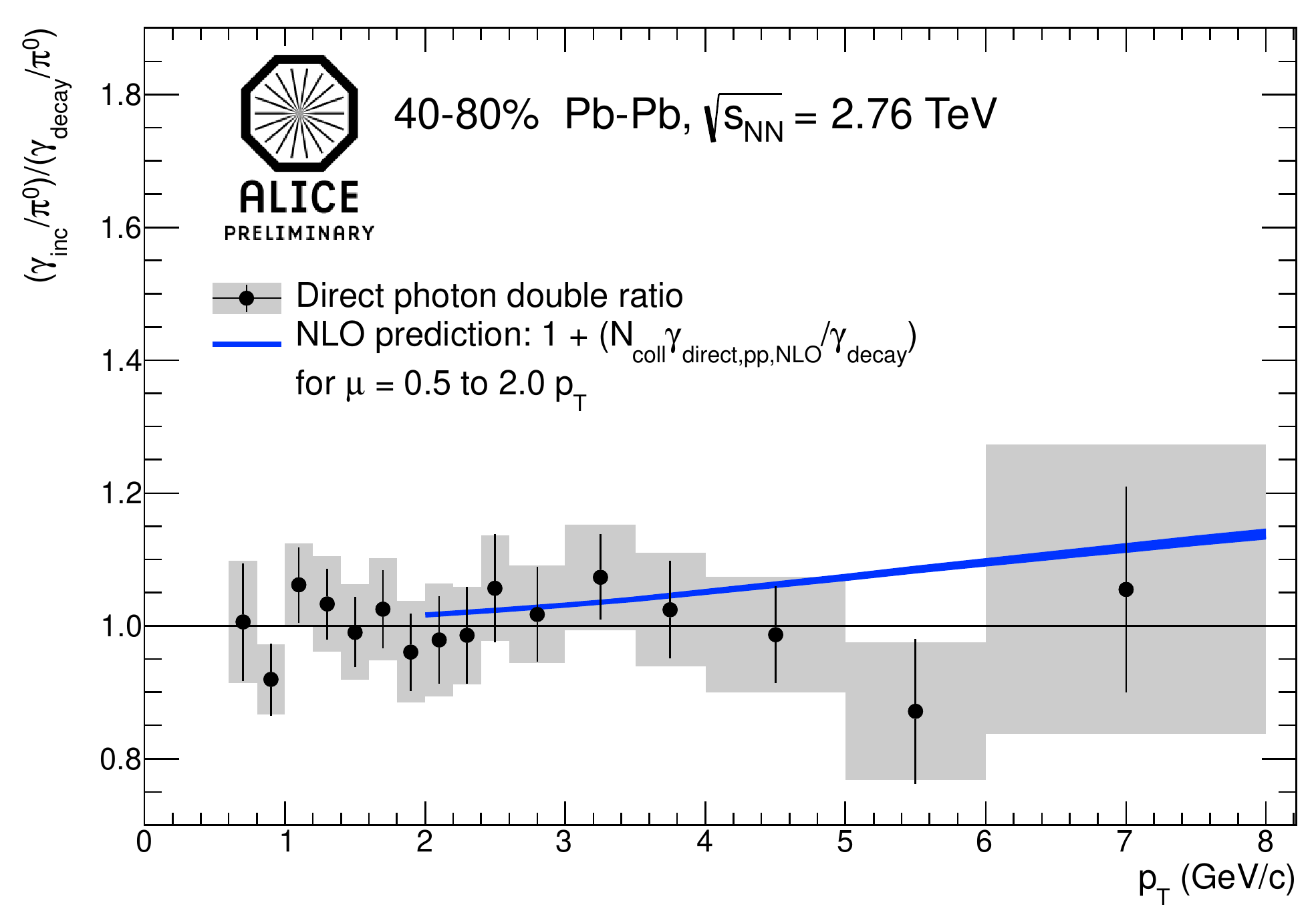}
\captionsetup{margin={5pt,0pt}}
\caption{(color online) Direct-photon double ratio in Pb-Pb collisions at $\sqrt{s_{_{NN}}}=\unit[2.76]{TeV}$ for 40-80\% centrality with NLO pQCD predictions}\label{convPbPbper}

\end{minipage}\hfill
\end{figure}

In Fig.\,\ref{doublepp} the double ratio (Eq.\,(\ref{eq:doubleratio})) for pp collisions at $\sqrt{s} = \unit[7]{TeV}$ is shown. Over the measured $p_{\mathrm{T}}$ range, the result is consistent with no direct-photon signal. The systematic uncertainties are too large to make a statement about the production of direct photons. A similar result for peripheral Pb-Pb collisions is presented in Fig.\,\ref{convPbPbper}. Both measurements are compared to NLO pQCD predictions for direct photons in pp collisions ($\gamma_{\mathrm{direct,NLO}}$ in Eq.\,\ref{eq:NLO}) \cite{Vogelsang}. These predictions are reformulated via
\begin{equation}
  \label{eq:NLO}
  \mathcal{R}_{\mathrm{NLO}}=1+\left( N_{\mathrm{coll}}\cdot\frac{\gamma_{\mathrm{direct,NLO}}}{\gamma_{\mathrm{decay}}}\right)
\end{equation}
to be comparable to a double ratio. $N_{\mathrm{coll}}$ is the average number of binary collisions from a Glauber Monte Carlo calculation (0-40\%: $N_{\mathrm{coll}} = 825$, 40-80\%: $N_{\mathrm{coll}} =78$) \cite{glauber}. In pp it is set to one. Both measurements, pp and peripheral Pb-Pb are in agreement with the NLO results, represented by a blue band.

\begin{figure}[h!]
\begin{minipage}[h]{0.5\textwidth}

\includegraphics[width=0.96\textwidth]{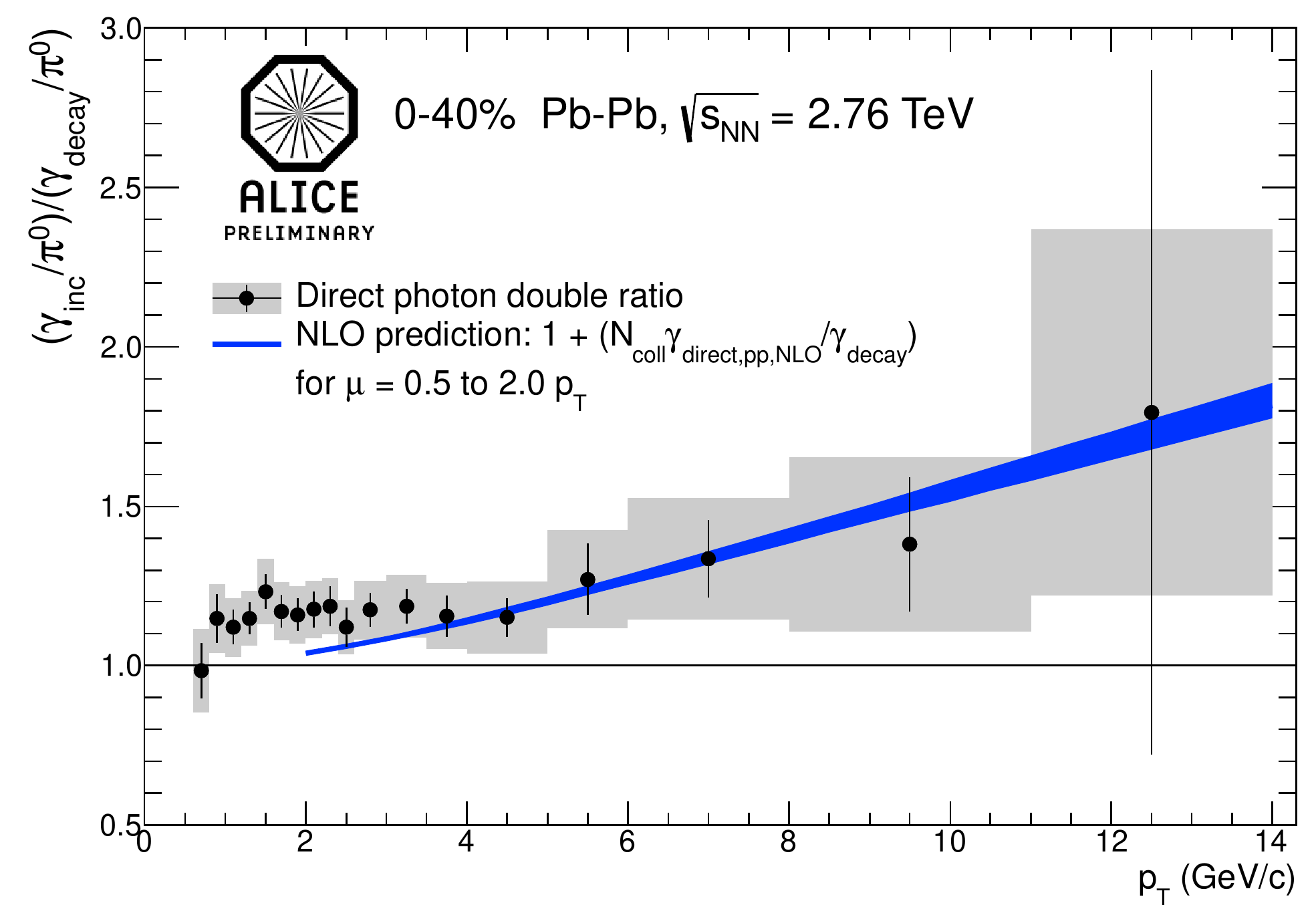}
\captionsetup{margin={0pt,5pt}}
\caption{(color online) Direct-photon double ratio in Pb-Pb collisions at $\sqrt{s_{_{NN}}}=\unit[2.76]{TeV}$ for 0-40\% centrality with NLO pQCD predictions}\label{convPbPbcent}

\end{minipage}\hfill
\begin{minipage}[h]{0.5\textwidth}

\includegraphics[width=0.96\textwidth]{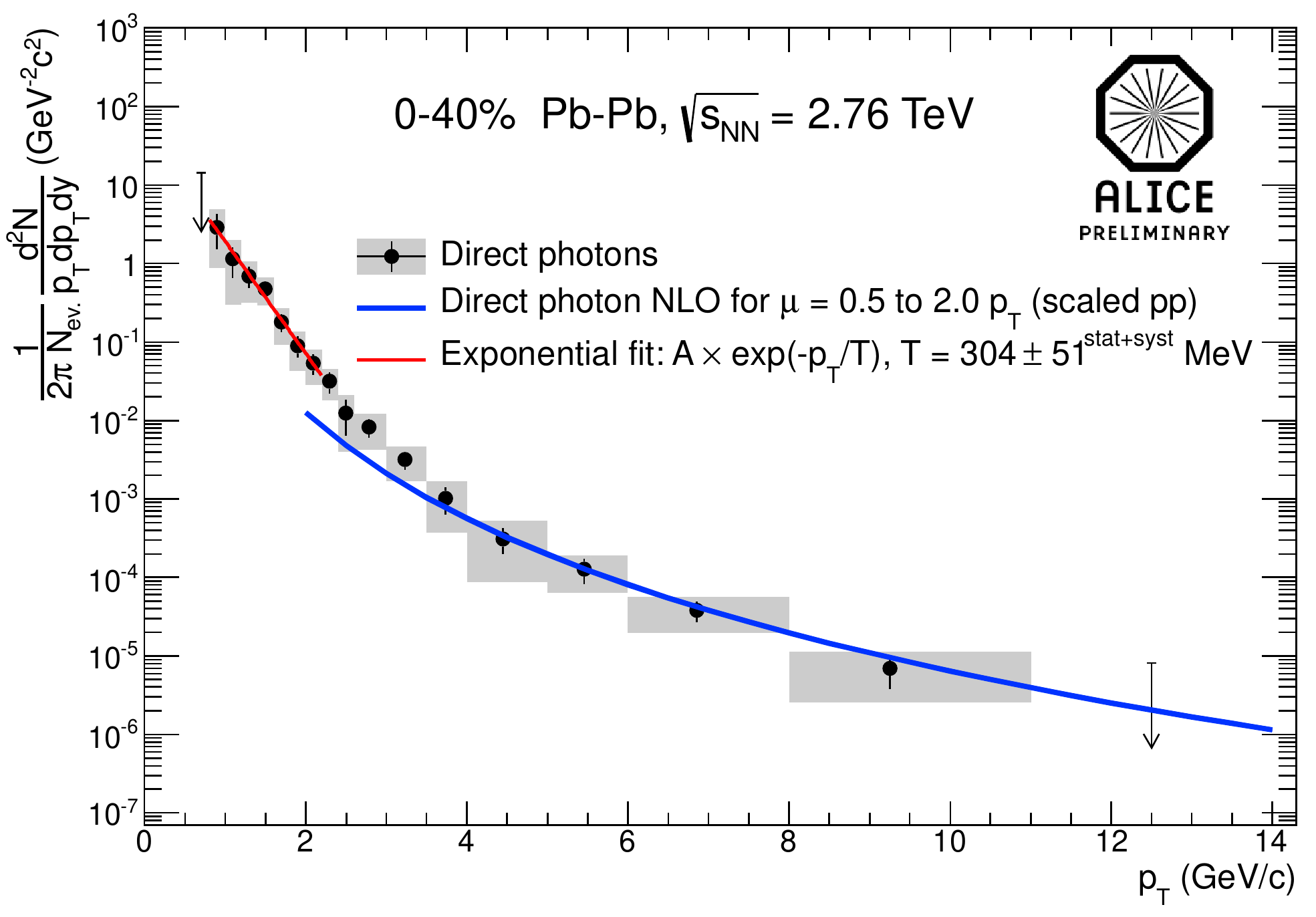}
\captionsetup{margin={5pt,0pt}}
\caption{(color online) Direct-photon invariant yield in Pb-Pb collisions at $\sqrt{s_{_{NN}}}=\unit[2.76]{TeV}$ for 0-40\% centrality with NLO pQCD predictions and exponential fit}\label{directspec}

\end{minipage}\hfill
\end{figure}

The situation changes in central Pb-Pb collisions, as shown in Fig.\,\ref{convPbPbcent}. The double ratio shows a clear signal over the whole range of $p_{\mathrm{T}}$.  Below $\unit[4]{GeV/\mathrm{c}}$ the direct-photon signal should contain a significant part of photons produced from a thermalized medium. Jet-photon conversions are also expected to contribute in this region.  From the double ratio the direct-photon yield is extracted using Eq.\,(\ref{eq:subs}) and shown in Fig.\,\ref{directspec}. Fig.\,\ref{directspec} also shows a direct-photon NLO calculation for pp at $\sqrt{s} = \unit[2.76]{TeV}$ scaled by $N_{\mathrm{coll}}$  \cite{Vogelsang} and an exponential fit to the low momentum part of the spectrum. The inverse slope parameter of the exponential for $\unit[0.8]{GeV/\mathrm{c}} < p_{\mathrm{T}} < \unit[2.2]{GeV/\mathrm{c}}$ is extracted as:
\begin{equation}\nonumber
  T_{\mathrm{LHC}}=\unit[304\pm 51^{\mathrm{syst+stat}}]{MeV}\nonumber.
\end{equation}\nonumber
In a similar analysis, PHENIX measures an inverse slope parameter of $T_{\mathrm{RHIC}} = \unit[221\pm 19^{\mathrm{stat}}\pm 19^{\mathrm{syst}}]{MeV}$ for 0-20\% Au-Au collisions at $\sqrt{s_{_{NN}}} = \unit[200]{GeV}$. In hydrodynamic models describing the PHENIX data, the inverse slope of \unit[220]{MeV} indicates an initial temperature of the QGP above the critical temperature $T_C$ for the transition to the QGP \cite{PHENIXResult1,PHENIXResult2}. The ALICE result shows an expected increase in the extracted temperature. This is the first measurement of a direct-photon signal at low $p_{\mathrm{T}}$ with real photons.

\section*{References}

\begin{thebibliography}{00} 



\bibitem{theorypp} Tomasz Pietrycki, Antoni Szczurek 2006 
  Int.J.Mod.Phys. A22 (2007) 541-545 (arXiv:hep-ph/0608190)
\bibitem{theoryPbPbpQCD1} Fu-Ming Liu, Klaus Werner 2009 
  J. Phys. G: Nucl. Part. Phys. 36 035101 (arXiv:0712.3619)
\bibitem{theoryPbPbpQCD2} Thomas Peitzmann, Markus H. Thoma 2002 
\bibitem{theoryPbPbpQCD3} Charles Gale 2009 Landolt-Boernstein Volume 1-23A (arXiv:0904.2184)
\bibitem{substraction} WA98 Collaboration 2000 (arXiv:nucl-ex/0006007)
\bibitem{substraction2} The STAR Collaboration 2009   Phys. Rev. C (arXiv:0912.3838)
\bibitem{ppPi0Paper} The ALICE Collaboration 2012 
  Phys. Lett. B 717, 162 (arXiv:1205.5724)
\bibitem{QM2012} Dimitry Peresunko, These Proceedings
\bibitem{phenixscaling} P. K. Khandai, P. Shukla, V. Singh 2011 
  Phys. Rev. C (arXiv:1110.3929)


\bibitem{Vogelsang} Werner Vogelsang 2010 Private Comunication
\bibitem{glauber} The ALICE Collaboration 2010 Phys.Rev.Let (arXiv:1011.3916)


\bibitem{PHENIXResult1} PHENIX Collaboration 2005 Phys.Rev.Lett.94:232301 (arXiv:nucl-ex/0503003)
\bibitem{PHENIXResult2} PHENIX Collaboration 2010 Phys.Rev.Lett.104:132301 (arXiv:0804.4168)


\end{thebibliography}
\end{document}